\documentclass[twocolumn]{aastex701}

\usepackage{multirow,booktabs}
\usepackage{enumitem}

\begin{document}
\title{ELUCID. IX. Recovering the Substructures and History of the Coma Cluster}

\author[0009-0006-0435-9469]{Xiong Luo} 
\affiliation{Key Laboratory for Research in Galaxies and Cosmology, Department of Astronomy, University of Science and Technology of China, Hefei, Anhui 230026, China}
\affiliation{School of Astronomy and Space Science, University of Science and Technology of China, Hefei 230026, China}
\email[show]{luoxiong@mail.ustc.edu.cn}

\author[0000-0002-4911-6990]{Huiyuan Wang} 
\affiliation{Key Laboratory for Research in Galaxies and Cosmology, Department of Astronomy, University of Science and Technology of China, Hefei, Anhui 230026, China}
\affiliation{School of Astronomy and Space Science, University of Science and Technology of China, Hefei 230026, China}
\email[show]{whywang@ustc.edu.cn}

\author[0000-0002-2113-4863]{Weiguang Cui} 
\affiliation{Departamento de F\'{i}sica Te\'{o}rica, Universidad Aut\'{o}noma de Madrid, M\'{o}dulo 15, E-28049 Madrid, Spain}
\affiliation{Centro de Investigaci\'{o}n Avanzada en F\'isica Fundamental (CIAFF), Facultad de Ciencias, Universidad Aut\'{o}noma de Madrid, 28049 Madrid, Spain}
\affiliation{Institute for Astronomy, University of Edinburgh, Royal Observatory, Edinburgh EH9 3HJ, United Kingdom}
\email{weiguang.cui@uam.es}

\author[0000-0002-4534-3125]{Yipeng Jing} 
\affiliation{Department of Astronomy, School of Physics and Astronomy, Shanghai Jiao Tong University, Shanghai 200240, China}
\affiliation{Tsung-Dao Lee Institute, Shanghai Jiao Tong University, Shanghai 200240, China}
\email{ypjing@sjtu.edu.cn}

\author[0000-0003-2204-2474]{Weipeng Lin} 
\affiliation{School of Physics and Astronomy, Sun Yat-sen University, DaXue Road 2, 519082, Zhuhai, China}
\affiliation{CSST Science Center for the Guangdong-Hongkong-Macau Greater Bay Area, DaXue Road 2, 519082, Zhuhai, China}
\email{linweip5@mail.sysu.edu.cn}

\author[0000-0002-3097-5381]{Neal Katz}
\affiliation{Department of Astronomy, University of Massachusetts, Amherst MA 01003-9305, USA}
\email{nsk@umass.edu}

\author[0000-0003-2842-9434]{Romeel Dav\'{e}}
\affiliation{Institute for Astronomy, University of Edinburgh, Royal Observatory, Edinburgh EH9 3HJ, United Kingdom}
\affiliation{University of the Western Cape, Bellville, Cape Town 7535, South Africa}
\affiliation{South African Astronomical Observatories, Observatory, Cape Town 7925, South Africa}
\email{Romeel.Dave@ed.ac.uk}

\author[0000-0001-5356-2419]{Houjun Mo} 
\affiliation{Department of Astronomy, University of Massachusetts, Amherst MA 01003-9305, USA}
\email{hjmo@astro.umass.edu}

\author[0000-0003-3997-4606]{Xiaohu Yang}
\affiliation{Department of Astronomy, School of Physics and Astronomy, Shanghai Jiao Tong University, Shanghai 200240, China}
\affiliation{Tsung-Dao Lee Institute, Shanghai Jiao Tong University, Shanghai 200240, China}
\affiliation{Shanghai Key Laboratory for Particle Physics and Cosmology, Shanghai Jiao Tong University, Shanghai 200240, China}
\email{xyang@sjtu.edu.cn}

\author[0000-0002-4326-3543]{Hao Li}
\affiliation{Key Laboratory for Research in Galaxies and Cosmology, Department of Astronomy, University of Science and Technology of China, Hefei, Anhui 230026, China}
\affiliation{School of Astronomy and Space Science, University of Science and Technology of China, Hefei 230026, China}
\email{lh123@mail.ustc.edu.cn}

\begin{abstract}
We employ constrained hydrodynamic simulations of the Coma galaxy cluster from the ELUCID project to study its substructures and assembly history.
Our simulations accurately reproduce the global properties of Coma, including its position, virial mass, radius, and surrounding large-scale filaments. 
Using a combined HBT+SKID method, we obtain a total intracluster light (ICL) fraction of $12.2\%-23.5\%$, consistent with recent observations.
The mass-weighted ICL profiles of velocity dispersion, stellar age, and iron abundance generally agree with MaNGA measurements, and its clumpy east-west elongation is closely linked to the merger history of the two brightest cluster galaxies (BCGs). 
The simulation also successfully reproduces the north and west intracluster filaments (ICFs) detected by weak lensing, and the surrounding galaxy groups are distributed in alignment with the directions toward the nearby A2199 and A1367 clusters. 
The simulation further predict a complex cluster assembly history, with two major mergers at $z=0.74$ and $0.45$, and a pericentric passage of the two BCGs occurring $\sim 0.64 \ {\rm Gyr}$ before its current state.
Our results demonstrate that constrained simulations are a powerful tool for connecting observed structures to the unobservable assembly histories of individual galaxy clusters.
\end{abstract}


\section{Introduction}
Galaxy clusters, as the largest gravitationally bound structures in the Universe, are the final products of hierarchical structure formation.
They form at the intersections of large-scale filaments where dark matter and baryons converge, and continue to grow through the accretion of mass from their surroundings and the merging of smaller substructures.
As the densest regions of the Universe, clusters are ideal laboratories for investigating the evolution of massive galaxies, the heating and cooling of intracluster medium (ICM), and the environmental effects on infalling satellite galaxies. 
Studying the detailed structures and assembly histories of nearby clusters provides a direct window into the physics of galaxy formation and the evolution of large‑scale structures \citep{Kravtsov2012ARA&A_Formation, Boselli2022A&ARv_RPS, Bourne2023Galax_radio_jet}.

Galaxy clusters, composed primarily of stars, gas, and dark matter, host a variety of substructures. 
These components and their substructures have been extensively studied through multi-wavelength observations, from radio to X-ray, particularly for nearby systems such as Virgo and Coma.
The stellar component encompasses a diverse range of objects, including the brightest cluster galaxy \citep[BCG,][]{Collins2003Ap&SS_BCG, Gerhard2007A&A_kinematics, Sanders2014MNRAS_Xray_2BCGs_Coma}, jellyfish galaxies \citep{Roberts2021A&A_LoTSS_jellyfish, Poggianti2025A&A_MUSE_RPS_GASP, Foster2026ApJ_CLIFS}, ultra‑diffuse galaxies \citep[UDGs,][]{Koda2015ApJ_UDG_Coma, Lim2018ApJ_UDG_Coma, Junais2022A&A_VESTIGE_UDG_Virgo, Bautista2023ApJS_UDG_HSC_Coma, Buttitta2025A&A_LEWIS_UDG_Hydra_I}, galaxy groups \citep{Adami2005A&A_infalling_substructures, Healy2021A&A_HI_Coma_substructures, Jimenez-Teja2025A&A_ICL_Coma}, stellar streams \citep{Roman2023A&A_stellar_stream_Coma}, and the diffuse intracluster light \citep[ICL,][]{Mihos2005ApJ_ICL_Virgo, Kluge2020ApJS_BCG_ICL, Arnaboldi2022FrASS_ICL, Montes2022NatAs_ICL, Jimenez-Teja2025A&A_ICL_Coma, Kluge2025A&A_Euclid_ICL_Perseus}.
The gas component includes the interstellar medium \citep[ISM,][]{Roberts2023arXiv231020417R_RPS_Coma, Edler2024A&A_ViCTORIA_Virgo, Broderick2025ApJ_RPS_Coma}, the ICM \citep{Matsushita2011A&A_ICM_XMM_Newton, Sato2011PASJ_Suzaku_ICM_Coma, Planck2013A&A_ICM_Coma, Mirakhor2020MNRAS_Coma, Gatuzz2025A&A_ICM_A3266}, and shock structures \citep{Markevitch2007PhR_shock, Uchida2016PASJ_shock_Coma, Churazov2023A&A_Coma_shock}.
Studies of the dark matter component focus on density profiles \citep{Kubo2007ApJ_WL_Coma, Umetsu2014ApJ_CLASH_WL}, subhalos \citep{Okabe2014ApJ_Subaru, Sasaki2015ApJ_Suzaku_Coma}, and intracluster filaments \citep[ICF,][]{HyeongHan2024NatAs_ICF, Shinde2025arXiv251026318S_WL_ICF}.
In parallel, the large‑scale environment in which clusters are embedded has also been the subject of active investigation \citep{Kim2016ApJ_LSS_Virgo, Malavasi2020A&A_spider}.

Nevertheless, observations alone cannot recover the full assembly history of galaxy clusters.
To overcome this limitation and gain deeper insight into their formation and evolution, previous studies have employed hydrodynamic simulations \citep[e.g.][]{Cui2018MNRAS_The_Three_Hundred, Zhang2021MNRAS_merging_cluster, Luo2024ApJ_ELUCIDVIII}.
In conventional simulations, the initial conditions are generated randomly, based on the initial density fluctuations predicted by cosmological models.
Therefore, meaningful comparisons between simulated and observed clusters require robust statistical analyses that demands a sufficiently large sample.
However, due to the low number density of clusters and the limitations of computational resources, the number of simulated clusters is ultimately constrained by the trade‑off between simulation box size and mass resolution.

There are two different approaches to increase the sample size of simulated clusters.
The first is to run hundreds of zoom‑in simulations, in which each simulation focuses on a single cluster and its immediate environment, while the outer regions are modelled with particles of lower resolution.
For example, simulation series such as TNG‑Cluster \citep{Nelson2024A&A_TNC_Cluster}, The Three Hundred \citep{Cui2018MNRAS_The_Three_Hundred}, Cluster‑EAGLE \citep{Barnes2017MNRAS_Cluster_EAGLE}, and DIANOGA \citep{Ragone-Figueroa2020MNRAS_DIANOGA} have simulated 352, 324, 30, and 29 clusters, respectively.
The second approach is to run a simulation with a sufficiently large box size to produce an adequate number of clusters, as exemplified by the FLAMINGO \citep{Schaye2023MNRAS_FLAMINGO} simulation series, whose largest simulation box reaches up to 2.8 Gpc.

These simulations have enabled a wide range of studies on cluster components such as the ICM and ICL.
However, these approaches still suffer from several limitations.
First, to ensure a statistically meaningful number of clusters, the computational cost required remains substantial.
Second, comparing simulations with observations is challenging due to the limited availability of high‑quality observational data for nearby clusters.
Finally, these studies often neglect the diversity among clusters, which arises from their different assembly histories and surrounding large-scale structures. 

An alternative approach is to use constrained simulations (or reconstruction simulations). 
Unlike conventional simulations, the initial condition of a constrained simulation is generated to reproduce the present-day density field of a specific region in the real Universe.
Consequently, a constrained simulation of an actual cluster allows us to directly compare simulation results with observational data for the individual cluster.
In this way, it eliminates the need to run a large ensemble of clusters and avoids the scatter introduced by cluster-to-cluster variations, making optimal use of the available high‑quality observational data for nearby clusters.

Several previous studies have used constrained simulations to investigate galaxy clusters.
For example, \citet{Olchanski2018A&A_Virgo} employed a simulation series based on initial conditions generated by the Constrained Local UniversE Simulations (CLUES) reconstruction method \citep{Sorce2016MNRAS_CLUES}.
By comparing the simulated Virgo clusters across all realizations with randomly selected halos, they concluded that, on average, Virgo-like halos have experienced only one merger with a mass larger than 10 percent of the final cluster mass within the last four Gyrs.
More recently, \citep{Sorce2021MNRAS_CLONE} introduced the Constrained LOcal and Nesting Environment (CLONE) simulation, a zoom-in hydrodynamical simulation centered on the Virgo cluster counterpart.
Their simulation reveals that approximately 300 small galaxies ($m_{*,\rm gal} > 10^7 \ {\rm M}_{\odot}$) entered the cluster within the last Gyr and the last significant merger event occurred about 2 Gyrs ago.

As one of the most massive clusters in the local Universe, the Coma cluster (A 1656) has also been studied with constrained simulations \citep{Dolag2016MNRAS_SZ, Jasche2019A&A_Bayesian, Boss2024A&A_SLOW_III, Groth2026ApJ_SLOW}.
For instance, \citet{Malavasi2023A&A_Coma_CS} applied the Discrete Persistent Structure Extractor \citep[DisPerSE,][]{Sousbie2011MNRAS_Disperse} to identify filaments using galaxies from both the constrained simulation and the real Coma cluster.
Their results show that the simulated Coma cluster successfully reproduces the north-east and west filaments, and that the number of connected filaments is consistent with the observations.
They further found that the accretion of matter onto the simulated Coma cluster is significantly more collimated near the filaments compared to the general isotropic accretion flow.

The aforementioned studies demonstrate that constrained simulations are a powerful tool for investigating the assembly history and structures of galaxy clusters.
In this work, we employ the constrained simulations of the Coma cluster presented in \citet{Luo2024ApJ_ELUCIDVIII} to compare with observational results and inferences regarding its various components and substructures.
Taking advantage of the high-fidelity replica of the Coma cluster, we further infer its assembly history and physical properties beyond the reach of current observations.

The remainder of this paper is organized as follows.
In \autoref{Sec:Data_and_Methods}, we describe the simulations, algorithms used to identify different structures and build merger trees, and the observational data employed in this work.
In \autoref{Sec:Comparisons}, we compare the general properties of the Coma cluster with observational results, along with its various components and substructures, including the two BCGs, ICL, ICF, and galaxy groups.
In \autoref{Sec:Predictions}, we present the predictions from the simulations, covering the assembly history of the Coma cluster, the evolution trajectories of the two BCGs, and the property distribution maps of the ICL.
Finally, in \autoref{Sec:Summary}, we summarize our main findings and provide further discussion.

\section{Data and Methods} \label{Sec:Data_and_Methods}
\subsection{Simulations}
We employ the constrained simulations of the Coma cluster presented in \citet{Luo2024ApJ_ELUCIDVIII}, which are part of the Jiutian simulation suite \citep{Han2025SCPMA_Jiutian}.
These simulations not only accurately reproduce the Coma cluster itself but also successfully capture the large-scale structures surrounding it. 
A brief description is provided below.

The initial conditions for these simulations are taken from the reconstruction of the local universe produced by the Exploring the Local Universe with reConstructed Initial Density field (ELUCID) project \citep{Wang2014ApJ_ELUCIDI, Wang2016ApJ_ELUCIDIII}.
We adopted the zoom-in technique to simulate the Coma cluster. The high-resolution area, a spherical region with a radius of at least $20 \ h^{-1}{\rm Mpc}$ centered on the Coma cluster, is contained within a simulation box with a side length of $500 \ h^{-1}{\rm Mpc}$.
The corresponding particle masses for the dark matter and initial gas particles in the high-resolution region are $3.20\times10^7 \ h^{-1}{\rm M}_{\odot}$ and $6.58\times10^6 \ h^{-1}{\rm M}_{\odot}$, respectively.
The adopted cosmological parameters are given by WMAP5 \citep{Dunkley2009ApJS_WMAP5}: $\Omega_{\Lambda} = 0.742$, $\Omega_{\rm m}=0.258$, $\Omega_{\rm b}=0.044$, $n_{\rm s}=0.96$, $\sigma_8=0.80$, $h=0.72$. 

The simulation set comprises five hydrodynamic simulations, each employing different subgrid physics models and simulation codes: G3-H-CM, GZ-SB-CM, GZ-SBnA-CM, GZ-SBrw-CM, GZ-SBrs-CM.
G3-H-CM is run using the GADGET-3 code and the galaxy formation model without AGN feedback from \citet{Huang2019MNRAS_robustness, Huang2020MNRAS_impact}.
GZ-SB-CM uses the same models as SIMBA \citep{Dave2019MNRAS_SIMBA}, evolved with the hydrodynamics code GIZMO \citep{Hopkins2015MNRAS_GIZMO}, while GZ-SBnA-CM is a version without black holes and AGN feedback.
GZ-SBrw-CM and GZ-SBrs-CM implement modified models based on the fiducial SIMBA framework to increase the total metal budget, with a different strategy in the jet mode AGN feedback between GZ-SBrw-CM and GZ-SBrs-CM \citep[see ][for more details]{Luo2024ApJ_ELUCIDVIII}.

In our previous work, we compared the results of various models with observations using a range of statistics, including the galaxy stellar mass functions, stellar mass--stellar metallicity relations, stellar mass--gas phase metallicity relations, and intracluster medium properties.
Since this study focuses on the substructures of the Coma cluster---which are more directly tied to the evolution of large-scale structures than to galaxy formation prescriptions---the differences among the models are minimal.
Therefore, unless otherwise stated, we present results primarily from the GZ-SBrw-CM simulation, which is overall closest to the observations among the available models.
This simulation provides 151 snapshots with redshifts ranging from 19 to 0. 
The simulated density field at redshift zero is designed to match the present-day mass density field derived from galaxies in the local universe, including the Coma cluster which has a redshift of 0.0241 \citep{Yang2007ApJ_Galaxy_Groups}.
Thus, when comparing with observations, we show simulation results mainly at redshift zero.

\subsection{Subhalos, Galaxies and Merger Trees}
To identify structures in the simulations, we first construct a halo catalogue using the Friends-of-Friends (FoF) algorithm.
All dark matter particles within a linking length of 0.2 times their mean separation are linked to each other.
Subsequently, gas, star, and black hole particles are linked to their nearest dark matter particle if the distance is less than the same linking length.
We define $M_{\rm FOF}$ as the total mass of all particles in a halo, and $M_{\rm *}$ as the mass of its stellar component.
Only halos with $M_{\rm FOF}$ exceeding 20 times the dark matter particle mass are retained.
The simulated Coma cluster is the most massive halo in the high-resolution region of the simulation.
The halo virial radius $R_{\rm 200c}$ is defined as the radius within which the mean enclosed density is 200 times the critical density, and $M_{\rm 200c}$ is the total mass within $R_{\rm 200c}$.

We then use the Hierarchical Bound-Tracing Plus algorithm \citep[hereafter HBT,][]{Han2012MNRAS_HBT, Han2018MNRAS_HBTPlus} to find subhalos and build their merger trees.
Using the FOF halo catalogue as input, HBT traces each halo from its birth, through its evolution as a satellite subhalo, until it is either destroyed or completely merges (i.e., loses all its particles to another subhalo).
Each FOF halo contains one central subhalo (the dominant one), with the rest being satellite subhalos if there are any.
The central subhalo includes the self-bound particles of the host halo after excluding all the satellite particles.
Each self-bound object is assigned a unique track ID at birth that remains unchanged throughout its entire lifetime, simplifying the tracking of its evolutionary history.
The tracking method preserves the hierarchical structure of satellite systems by recording the track IDs of sub-subhalos within a satellite subhalo from the time before it became a satellite, enabling the identification of subgroups in a galaxy cluster.
The center of a subhalo is defined as the position of its most bound particle, and the center of a halo is defined as the center of its central subhalo.

To investigate and track individual galaxies in the simulations, we also identify and build merger trees for galaxies.
We use the Spline Kernel Interpolative Denmax algorithm \citep[SKID,][]{Keres2005MNRAS_SKID} to find galaxies based on the density field, the same catalogue used in our previous study \citep{Luo2024ApJ_ELUCIDVIII}.
Each galaxy includes all the self-bound star particles and star-forming gas particles that connect to the same local density peak, which is defined as the galaxy center.
The galaxy stellar mass ($m_{*,\rm gal}$) is the total mass of its star particles.
The galaxy velocity is computed as the mass-weighted mean velocity of all member star particles.
Galaxies with at least 8 member particles are retained.
To trace their evolutionary histories, we construct merger trees in which the main progenitor of a given galaxy is defined as the most massive galaxy in the previous snapshot that has at least half of its stellar mass transferred to the descendant.
We also match each galaxy to its host subhalo identified by HBT, if the subhalo shares at least half of the galaxy's stellar mass.

\subsection{The Two Brightest Cluster Galaxies and the Intracluster Light}

\begin{figure*}[htbp!]
    \centering
    \includegraphics[width=16cm]{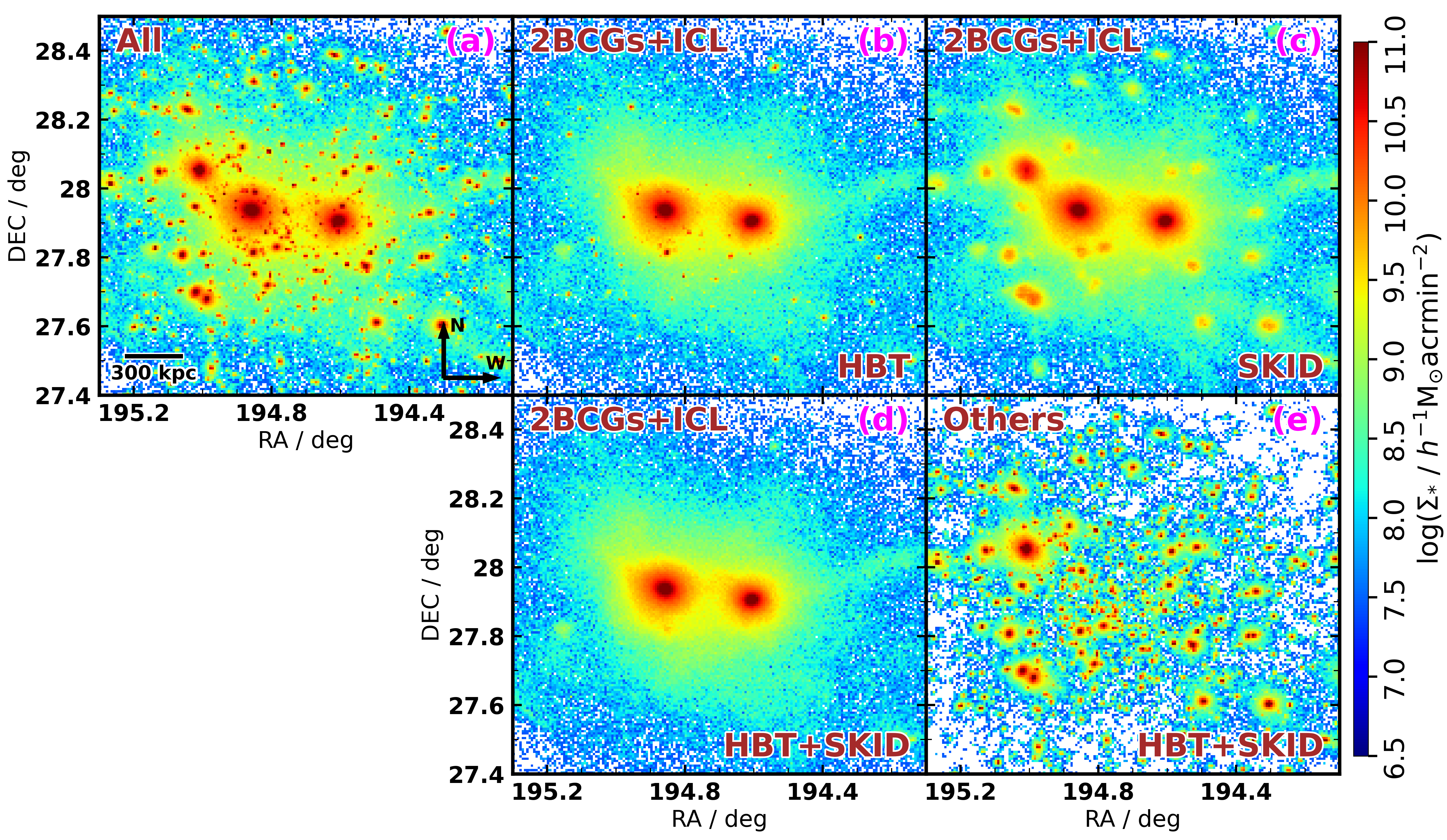}
    \caption{Stellar surface density maps at redshift zero viewed from the earth in the GZ-SBrw-CM simulation, projected within a line-of-sight depth of $\pm 5 \ h^{-1}{\rm Mpc}$ centered on the Coma cluster. 
    The axes show Right Ascension (RA) and Declination (DEC) in J2000.0 coordinates, with arrows indicating north and west in the lower-right corner of panel (a).
    The different panels correspond to different stellar components: (a) all star particles; (b) 2BCGs+ICL identified by HBT; (c) 2BCGs+ICL identified by SKID; and (d) and (e) 2BCGs+ICL and all the excluded particles (Others) identified by HBT+SKID, respectively.}
    \label{fig:2BCGsICL_definition}
\end{figure*}

The Coma cluster contains two BCGs: NGC 4889 (the eastern one) and NGC 4874 (the western one).
As shown in panel (a) of \autoref{fig:2BCGsICL_definition}, the two BCGs are clearly reproduced in our constrained simulation of the Coma cluster.
The stellar surface density (projected on the sky, with west to the right and north up) reveals two massive galaxies near the cluster center, mirroring the real Coma.
We refer to the eastern and western BCGs as BCG‑E and BCG‑W, respectively.
Owing to the unclear boundary between the BCG and ICL, earlier studies commonly modeled them together as a single BCG+ICL component \citep{Brough2024MNRAS_ICL_fractions, Kimmig2025A&A_ICL}. 
In the case of the Coma cluster, we adopt the notation 2BCGs+ICL to denote the combined contribution of both BCGs and the ICL.

To define the 2BCGs+ICL component, we first tested two approaches: one excludes all star particles in all subhalos except those in the host subhalos of the two BCGs, using HBT-identified subhalos; the other does the same but using galaxies found by SKID. 
The corresponding projected stellar surface density maps, labeled ``HBT'' and ``SKID'', are shown in panels (b) and (c) of \autoref{fig:2BCGsICL_definition}, respectively.
As one can see, the HBT method effectively removes satellite subhalos, particularly the larger ones.
However, some small stellar clumps near the BCGs remain.
In contrast, the SKID method eliminates all stars located in dense regions but retains the stellar bubble shells, which stay attached to their host galaxies.
Neither method is capable of producing a well-defined and smooth 2BCGs+ICL component.

The result of the SKID method is similar to conventional ICL definitions, such as removing all star particles bound to satellite galaxies found by other algorithms or excluding those within a radial distance of 30 kpc from subhalos.
These conventional approaches are thought to overestimate the ICL due to the inclusion of the galactic outskirts that remain kinematically aligned with the satellite galaxies, as discussed in \citet{Jeon2026ApJ_NEWCLUSTER}.
Instead, \citet{Jeon2026ApJ_NEWCLUSTER} used merger trees to remove both ``loosely bound'' and ``strictly bound'' particles of satellites.
Given that HBT traces subhalos while building merger trees, it is well suited for defining the ICL and can yield a BCG+ICL result similar to their method after removing satellite subhalos identified by HBT.
However, HBT fails to trace ejected satellites that once completely merged into another subhalo (particularly the central subhalo) but remain within the same host FOF halo.
Galaxy finders based on the density field, such as SKID, can complement this approach by identifying such satellites, whose outskirts are already completely disrupted.

We combine the two methods by removing all star particles in satellite subhalos identified by HBT or in satellite galaxies identified by SKID.
In practice, we first select all star particles belonging to the two subhalos (including the central subhalo of Coma) identified by HBT that host the two BCGs.
The 2BCGs+ICL component is then defined by removing any particles belonging to SKID-identified galaxies other than the two BCGs.
The bottom row of \autoref{fig:2BCGsICL_definition} shows the resulting stellar decomposition using this combined HBT+SKID approach.
Compared with the original methods shown in panels (b) and (c), the new method in panel (d) yields a significantly clearer and purer 2BCGs+ICL component.
Panel (e) displays all the excluded particles.

\subsection{Observations}
We also adopt the galaxy properties from the New York University Value-Added Galaxy Catalog \citep{Blanton2005AJ_NYUVAC}, based on the Sloan Digital Sky Survey Data Release 7 (SDSS DR7), together with the corresponding group catalogue from \citet{Yang2007ApJ_Galaxy_Groups}.
The provided Right Ascension (RA), Declination (DEC), and redshift for galaxies and the Coma cluster, are taken from these catalogues.
Galaxy stellar masses are derived from the relation between stellar mass-to-light ratio and color, using model magnitudes \citep[][]{Bell2003ApJS_Optical, Yang2007ApJ_Galaxy_Groups}.
These data serve as input for the original ELUCID reconstruction.

The ICL properties of the Coma cluster have also been measured observationally by \citet{Gu2020ApJ_ICL_Coma}.
They used spectroscopic data from three ICL regions in the Mapping Nearby Galaxies at Apache Point Observatory (MaNGA) Coma Deep program and stacked the 127 individual fiber spectra within each integral field unit (IFU).
From these data, they derived the velocity dispersion, stellar age, and iron abundance profiles of the ICL.
They presented these profiles as functions of the projected angular distance to the nearest BCG, defined as the minimum of the two distances to each BCG.
This choice accounts for the fact that ICL regions around the two BCGs may be dynamically connected to different BCGs.
Following this approach, we also define the projected angular distance to BCGs ($\theta_{\rm p}$) in our simulations to compute the corresponding profiles.
The corresponding projected physical distance to the nearest BCG is denoted as $r_{\rm p}$.
Our resulting profiles are similar to those centered on either BCG individually, but our definition reduces the anisotropy caused by the presence of the two BCGs and provides a more physically meaningful description for the Coma cluster.
For the iron abundance $\rm [Fe/H]=\log({N}_{Fe}/{N}_{H})-\log({N}_{Fe}/{N}_{H})_{\odot}$ in the simulations, we adopt the abundance ratio of iron to hydrogen, $\rm \log({N}_{Fe}/{N}_{H})_{\odot}=-4.33$, from \citet{Anders1989GeCoA_Abundances}.

\section{Comparisons with Observations} \label{Sec:Comparisons}

In this section, we compare our simulated Coma cluster to the observed one. We concentrate on the internal substructures of the Coma cluster, including the two BCGs, the ICL, the ICF, and the surrounding galaxy groups. 
These substructures are in highly non-linear regions and are therefore expected to be extremely challenging for any reconstruction method to recover accurately. 
This comparison allows us to assess both the robustness and the capabilities of our constrained simulation.

\subsection{Global properties of the Coma cluster}
The simulated Coma cluster is located at $\rm RA=194.86^{\circ}, \ DEC=27.94^{\circ}$, with a distance of $\rm D=99.6 \ {\rm Mpc}$ at redshift zero.
This is in excellent agreement with the real Coma cluster, which lies at $\rm RA=194.81^{\circ}, \ DEC=27.95^{\circ}$ and $\rm D = 99.9 \ {\rm Mpc}$, as reported in the group catalogue of \citet{Yang2007ApJ_Galaxy_Groups}.
For comparison, other constrained simulations of the local universe, such as SIBELIUS-DARK \citep{McAlpine2022MNRAS_SIBELIUS-DARK}, give $\rm RA=196.76^{\circ}, \ DEC=30.13^{\circ}, D=108.2 \ {\rm Mpc}$, while \citet{Jasche2019A&A_Bayesian} give $\rm RA=195.76^{\circ}, \ DEC=28.15^{\circ}$, and $\rm D=88.9 \ Mpc$.

The simulated Coma cluster has a halo virial mass of $M_{\rm 200c}=7.3\times10^{14} \ h^{-1}{\rm M}_\odot$, consistent with the real Coma cluster, for which weak lensing measurements give $M_{\rm 200c}=6.2\times10^{14} \ h^{-1}{\rm M}_\odot$ \citep{Okabe2014ApJ_Subaru}, in line with results from other methods \citep[see][ and references therein]{Ho2022NatAs_dynamical_mass_Coma}.
The corresponding virial radius is $R_{\rm 200c} = 1.47 \ h^{-1}{\rm Mpc}$, which translates to an angular size of $\theta_{\rm 200c} = 70.4'$, in excellent agreement with the observational value of $70'$ derived from X-ray observations \citep{Simionescu2013ApJ_Thermodynamics}, and $2.06^{+0.07}_{-0.05} \ {\rm Mpc}$ fitted using red sequence galaxies from DESI \citep{Pedratti2026arXiv260316706P_DESI}.

Moreover, the constrained simulation can also reproduce the surrounding large-scale filaments along with the A 1367 cluster close to the Coma cluster.
Using the Discrete Persistent Structure Extractor (DisPerSE) algorithm on the SDSS DR7 Main Galaxy Sample, \citet{Malavasi2020A&A_spider} identified three secure filaments connected to the Coma cluster, pointing towards the NE, N, and W.
The filaments in the NE and W directions are clearly visible in the constrained simulations, as demonstrated in our previous works \citep{Li2022ApJ_ELUCIDVII, Luo2024ApJ_ELUCIDVIII}.
However, the filament toward the N is not clear, which is identified in their work as the least secure among the first-generation filaments, lacking any clear X-ray or SZ features.

\subsection{The Two Brightest Cluster Galaxies} \label{SubSec:BCGs}

There are two BCGs residing in the Coma cluster: NGC 4889 (the eastern BCG) and NGC 4874 (the western BCG).
Their redshifts are 0.0215 and 0.0239, respectively, corresponding to a line-of-sight (LoS) velocity difference of $\Delta v_{\rm LoS} = 722 \ {\rm km}/{\rm s}$ for NGC 4874 relative to NGC 4889.
The angular separation between them is $\theta_{\rm 2BCGs} = 7.2'$, and the position angle of NGC 4889 relative to NGC 4874 is $\varphi_{\rm 2BCGs} = 98.4^{\circ}$ measured as the angle east of north.

\begin{figure*}[htbp!]
    \centering
    \includegraphics[width=16cm]{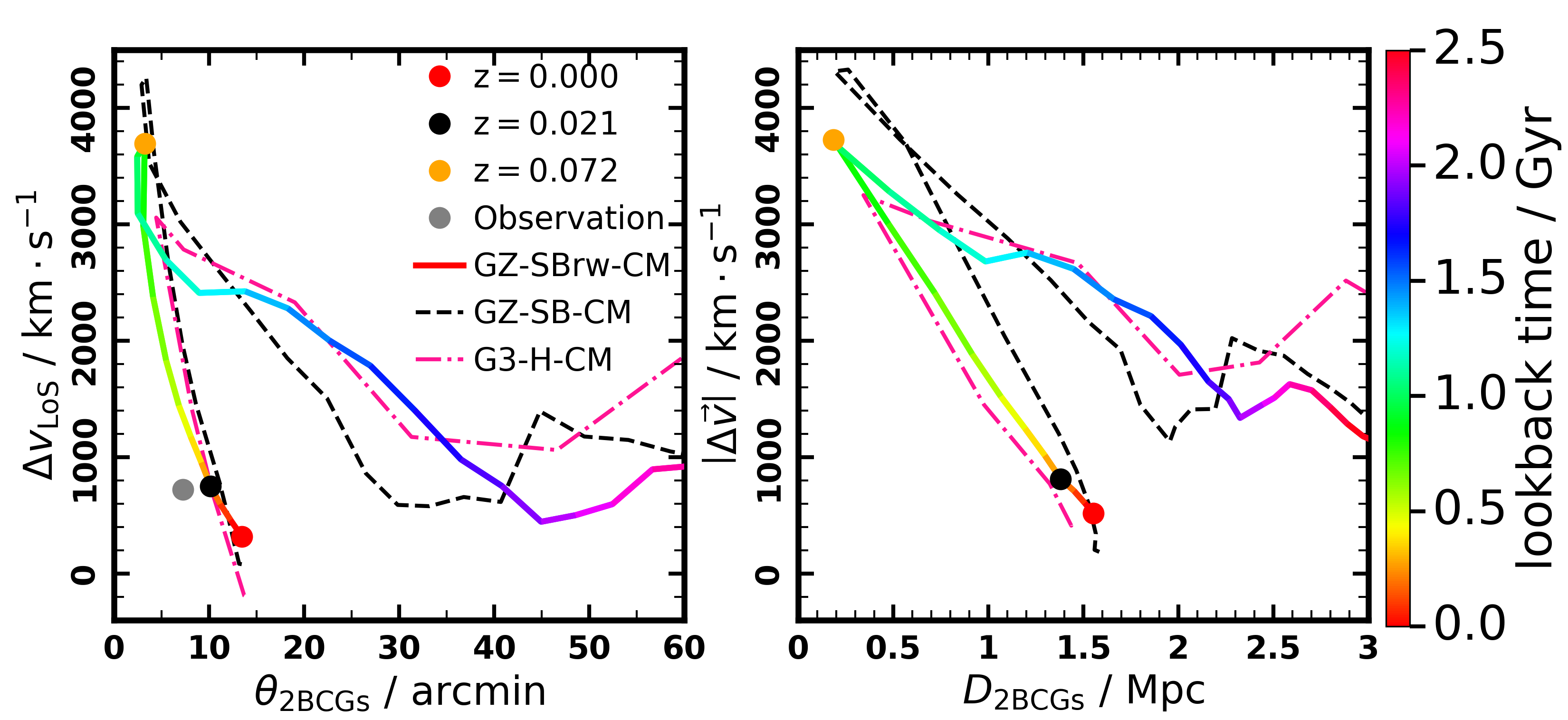}
    \caption{
    Evolution of the distance and relative velocity between the two BCGs in the simulation.
    Left panel: projected angular distance in the sky plane ($\theta_{\rm 2BCGs}$) versus line-of-sight velocity of BCG-W relative to BCG-E ($\Delta v_{\rm LoS}$).
    Right panel: similar to left but for their 3-D distance ($D_{\rm 2BCGs}$) and relative velocity ($\left| \Delta \vec{v} \right|$).
    The solid lines show the temporal evolution in GZ-SBrw-CM, color-coded by lookback time.
    The red, black, and orange points in each panel highlights the simulation results at $z=0$, $z=0.021$, and $z=0.072$, respectively.
    The gray point in the left panel indicates the observed value for the real Coma cluster.
    For comparison, the results for GZ-SB-CM and G3-H-CM are shown by the black dashed and pink dash-dotted lines, respectively.
    }
    \label{fig:2BCGs_D_V}
\end{figure*}

We first compare the separation and velocity difference between the two simulated BCGs with the observation.
In the left panel of \autoref{fig:2BCGs_D_V}, the solid colored line shows the temporal evolution of their projected angular distance ($\theta_{\rm 2BCGs}$) versus the LoS velocity of BCG-W relative to BCG-E ($\Delta v_{\rm LoS}$).
At redshift zero (red point), the simulation yields $\Delta v_{\rm LoS} = 319 \ {\rm km}/{\rm s}$, $\theta_{\rm 2BCGs} = 13.5'$, and $\varphi_{\rm 2BCGs} = 81.9^{\circ}$ (position angle of BCG-E relative to BCG-W).
Interestingly, at redshift 0.021 (black point), the simulation gives values $\Delta v_{\rm LoS} = 751 \ {\rm km}/{\rm s}$, $\theta_{\rm 2BCGs}=10.2'$, and $\varphi_{\rm 2BCGs} = 87.9^{\circ}$, which are much closer to the observed values.
Simulations from other models, GZ-SB-CM (dashed black line) and G3-H-CM (pink dash-dotted line), show the same trend.
This implies that, in terms of the two BCGs, our simulated cluster more closely matches the observed Coma cluster at a slightly higher redshift, corresponding to a time offset of $\sim 0.28 \ {\rm Gyr}$.
A similar trend was reported in our earlier analysis of the ICM profiles in \citet{Luo2024ApJ_ELUCIDVIII}. 
This time offset is not surprising given the highly nonlinear nature of these central regions.

\begin{figure*}[htbp!]
    \centering
    \includegraphics[width=16cm]{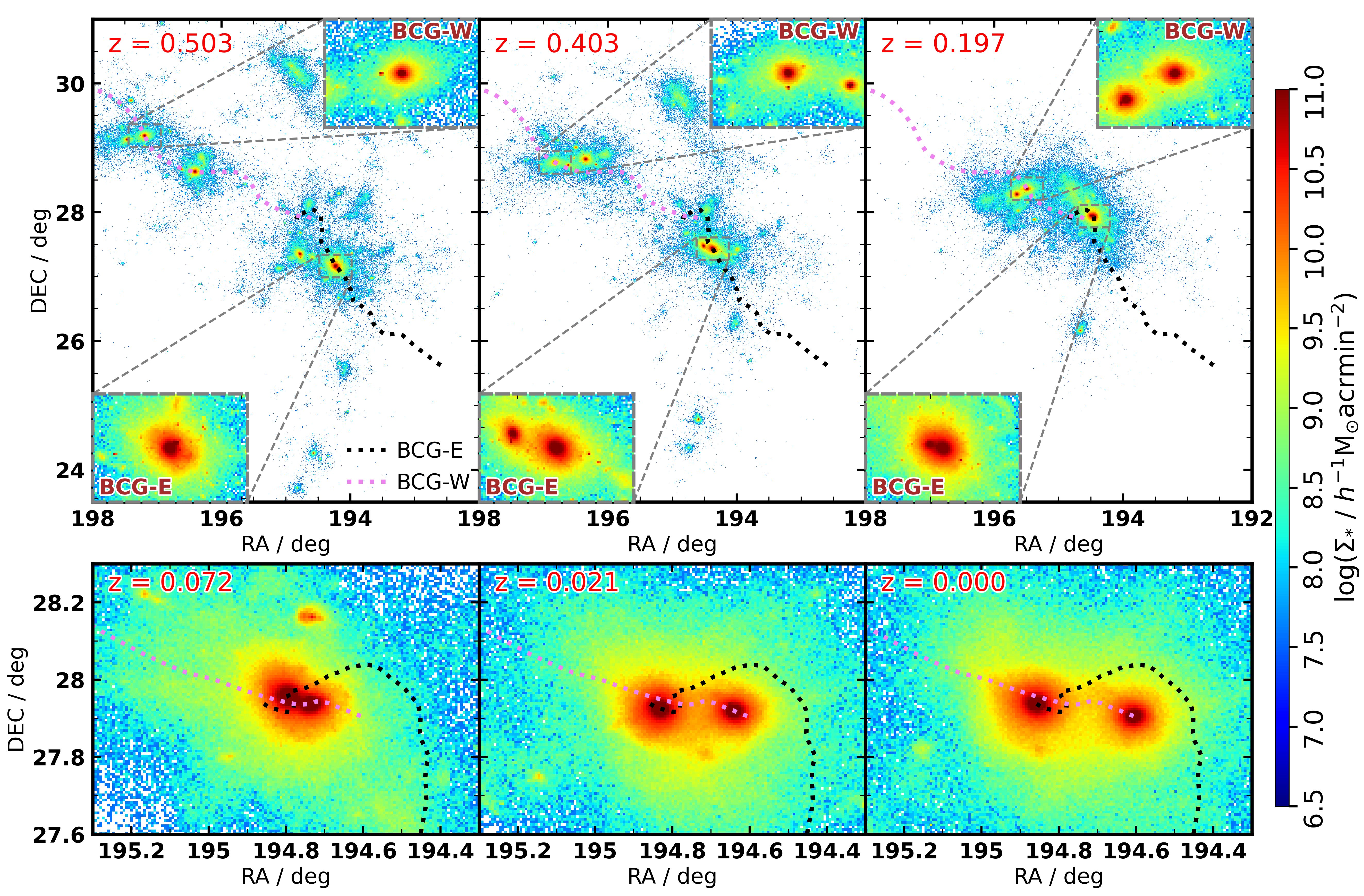}
    \caption{
    Evolution of the star particles in the 2BCGs+ICL component identified at redshift zero.
    The various panels display projected surface density maps at several redshifts, constructed by tracking all particles that make up the Coma cluster’s 2BCGs+ICL component at $z=0$ (shown in the lower-right panel). 
    The redshift of each panel is indicated in its upper-left corner. 
    In top panels, the lower-left and upper-right insets provide zoom-in views of the region around the two BCGs at the corresponding time, displayed at the same scale as the bottom panels.
    The black and pink dotted lines trace the trajectories of BCG-E and BCG-W, respectively. 
    }
    \label{fig:2BCGsICL_2Dmap}
\end{figure*}

\citet{Gerhard2007A&A_kinematics} analyzed the line-of-sight velocities of 37 intracluster planetary nebulae (ICPNe) associated with the ICL at about $5'$ south of NGC 4874.
They concluded that the NGC 4889 subcluster likely entered the cluster along the eastern A 2199 filament, in a direction nearly in the plane of the sky, met the NGC 4874 subcluster arriving from the west and has already passed through it.
\autoref{fig:2BCGsICL_2Dmap} shows the trajectories of the two BCGs in the simulation.
The black and pink dotted lines represent the tracks of BCG-E and BCG-W in the plane of the sky, respectively.
In line with the scenario inferred from the observations, the two BCGs appear to have already passed through pericenter, with one entering from the eastern filament and the other from the western side (see the lower-right panel).
In the simulation, however, their roles are interchanged: BCG-W is instead the one infalling from the east.
This mismatch may stem from the challenges of non-linear reconstruction, especially in capturing the detailed dynamics within the cluster core.
Even so, the simulation results remain plausible, as the observations do not directly constrain the complete dynamical history.
The simulation further suggests that, on larger scales, the two BCGs approach from directions roughly corresponding to the northeast and southwest, as illustrated in the upper panels.

\citet{Gerhard2007A&A_kinematics} also suggested that the two BCGs may converge in the plane of the sky, based on their low relative line-of-sight velocity and the unchanged, unshifted line-of-sight velocity distributions of the surrounding galaxies.
In contrast, our simulation offers an alternative picture: the two BCGs approach each other close to the line-of-sight direction, pass by each other, and have since reached a line-of-sight separation of more than $1 \ {\rm Mpc}$ and are now near apocenter, resulting in a low relative line-of-sight velocity (see \autoref{fig:2BCGs_D_V}).
Both possible scenarios were previously discussed in \citet{Fitchett1987ApJ_Substructure}.

\subsection{Intracluster Light}
\subsubsection{ICL Fraction} \label{SubSubSec:ICL_fraction}
The ICL fraction of the real Coma cluster is reported to be $\sim7\%-21\%$ by \citet{Jimenez-Teja2019A&A_J-PLUS_ICL_Coma}, depending on the optical wavelength.
More recently, \citet{Jimenez-Teja2025A&A_ICL_Coma} find $19.9\pm0.5\%$ in the $g$ band and $19.6\pm0.6\%$ in the $r$ band, with corresponding maximum detection radii of $1.33 \ \rm Mpc$ and $1.59 \ \rm Mpc$, respectively.
Based on the CHEFs Intracluster Light Estimator (CICLE), they removes the light from all the galaxies that lie in the field of view by fitting models with Chebyshev rational functions and Fourier series (CHEFs).
To compare the ICL fraction with their results, we define the ICL in the simulation as the 2BCGs+ICL component excluding the two BCGs identified by SKID.
The total ICL mass fraction, $M_{*, \rm ICL}/M_{*}$, is then computed as the ratio of the total mass of ICL particles to that of all star particles in the cluster.
The cumulative fraction profile, $M_{*, \rm ICL}(<\theta_{\rm p})/M_{*}(<\theta_{\rm p})$, represents the ratio of enclosed masses within a given projected angular radius. 
The corresponding mass fraction profile, $\Sigma_{*, \rm ICL}(\theta_{\rm p})/\Sigma_{*}(\theta_{\rm p})$, is the ratio of surface densities of the ICL and the total stellar component at each projected radius.
We also compute the same fraction profiles for the 2BCGs and the 2BCGs+ICL components.

\begin{figure*}[htbp!]
    \centering
    \includegraphics[width=16cm]{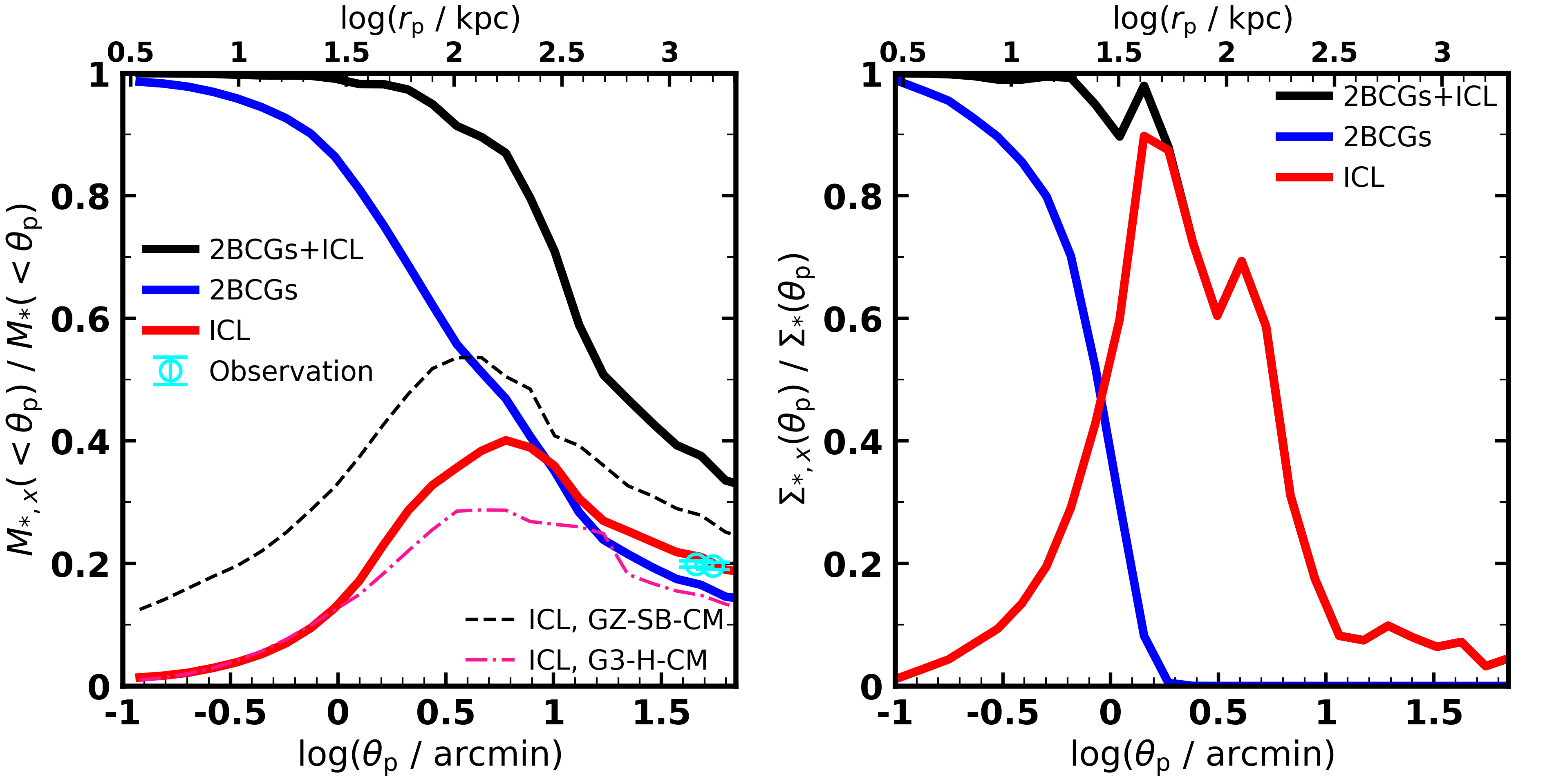}
    \caption{The cumulative mass fraction profiles (left panel) and surface mass density ratio profiles (right panel) for different stellar components of the Coma cluster at $z=0$.
    The horizontal axis shows the projected angular distance ($\theta_{\rm p}$) to the nearest BCG, while the corresponding projected physical distance ($r_{\rm p}$) is given on the top axis.
    In the left panel, the solid lines represent the cumulative mass fraction within $\theta_{\rm p}$ for different components in GZ-SBrw-CM, i.e., $M_{*, x}(<\theta_{\rm p})/M_{*}(<\theta_{\rm p})$, where x denotes the component: 2BCGs+ICL (black), 2BCGs (blue), and ICL (red).
    The black dashed and pink dash-dotted lines show the results for the ICL in GZ-SB-CM and G3-H-CM, respectively. 
    Observational data from \citet{Jimenez-Teja2025A&A_ICL_Coma} are plotted as cyan circles with error bars.
    In the right panel, the lines show the surface mass density ratio profiles $\Sigma_{*, x}(\theta_{\rm p})/\Sigma_{*}(\theta_{\rm p})$, i.e., the mass fraction at each radius.
    }
    \label{fig:ICL_fraction_profile}
\end{figure*}

\autoref{fig:ICL_fraction_profile} shows the cumulative mass fraction within different projected angular radii (left panel) for various stellar components, along with the mass fraction profiles (right panel).
Solid lines of different colors represent different components, as indicated in the legend.
The total ICL mass fraction of the Coma cluster is $18.3\%$, while the enclosed fraction (solid red line) at radii of $1.33 \ \rm Mpc$ and $1.59 \ \rm Mpc$ is about $21.2\%$ and $20.0\%$, respectively, consistent with the observational results in \citet{Jimenez-Teja2025A&A_ICL_Coma} (cyan data points).
Note that relaxed clusters generally exhibit ICL fractions below 10\% \citep[see Figure 6 in][]{Jimenez-Teja2025A&A_ICL_Coma}.
This suggests a connection between the ICL fraction and the dynamical state of the cluster. 
Consequently, the agreement between our simulation and the observations implies that our reconstruction accurately captures the dynamical state of the Coma cluster. 
The black dashed and pink dash-dotted lines show the results for the other models, which yield total ICL fractions ranging from $12.2\%$ (G3-H-CM) to $23.5\%$ (GZ-SB-CM). 
This clearly indicates that the ICL fraction is sensitive to the specific galaxy formation model adopted.

The total 2BCGs+ICL fraction for GZ-SBrw-CM is $32.1\%$, and ranges from $26.2\%$ to $33.3\%$ for GZ-SB-CM and G3-H-CM.
These values agree with the BCG+ICL fractions of $\sim10\%-40\%$ reported for low-redshift clusters with comparable halo masses \citep{Montes2022NatAs_ICL}. 
Within a projected radius of $r_{\rm p} \sim 500 \ {\rm kpc}$, the 2BCGs+ICL component accounts for half of the total stellar mass.
We emphasize that these values depend on how one defines the ICL in observations and simulations \citep{Cui2014MNRAS_ICL, Montenegro-Taborda2025MNRAS_Photometric_TNG300}, as well as on the specific implementations of the simulation models themselves \citep{Kimmig2025A&A_ICL}.
Further in-depth investigations are therefore necessary.

\subsubsection{Surface Density Map}
Unlike those of relaxed clusters, the ICL in the Coma cluster is anisotropic even near the center. 
The ICL in the core of the Coma cluster is observed to be clumpy and elongated in the east-west direction, which is thought to be associated with the dynamical history of the two BCGs \citep{Thuan1977PASP_Photographic, Korchagin2001ApJ_Origin, Adami2005A&A_diffuse_light, Jimenez-Teja2019A&A_J-PLUS_ICL_Coma, Jimenez-Teja2025A&A_ICL_Coma}.

The color maps in \autoref{fig:2BCGsICL_2Dmap} present the projected distribution of all 2BCGs+ICL particles at $z=0$ in the simulation, traced back to earlier cosmic times with the corresponding redshifts indicated in the upper-left corner of each panel.
The top row shows several snapshots before the two BCGs approach each other, with insets displaying zoomed-in views of each BCG.
After their pericenter passage at $z=0.072$ (lower-left panel), the clumpy ICL structure is also reproduced around the two BCGs, elongated along the line connecting them (i.e., the east–west direction).
This feature is visible both at $z=0$ (lower-right panel) and at $z=0.021$ (lower-middle panel), the snapshot where the relative projected distance and line-of-sight velocity of the two BCGs most closely match the observed values.
Further studies are needed to investigate the connections between the merger history of the two BCGs, the stripping of subhalo outskirts, the disruption of dwarf galaxies, and the formation history of the ICL clump.

\subsubsection{Property Profiles}
\begin{figure*}[htbp!]
    \centering
    \includegraphics[width=16cm]{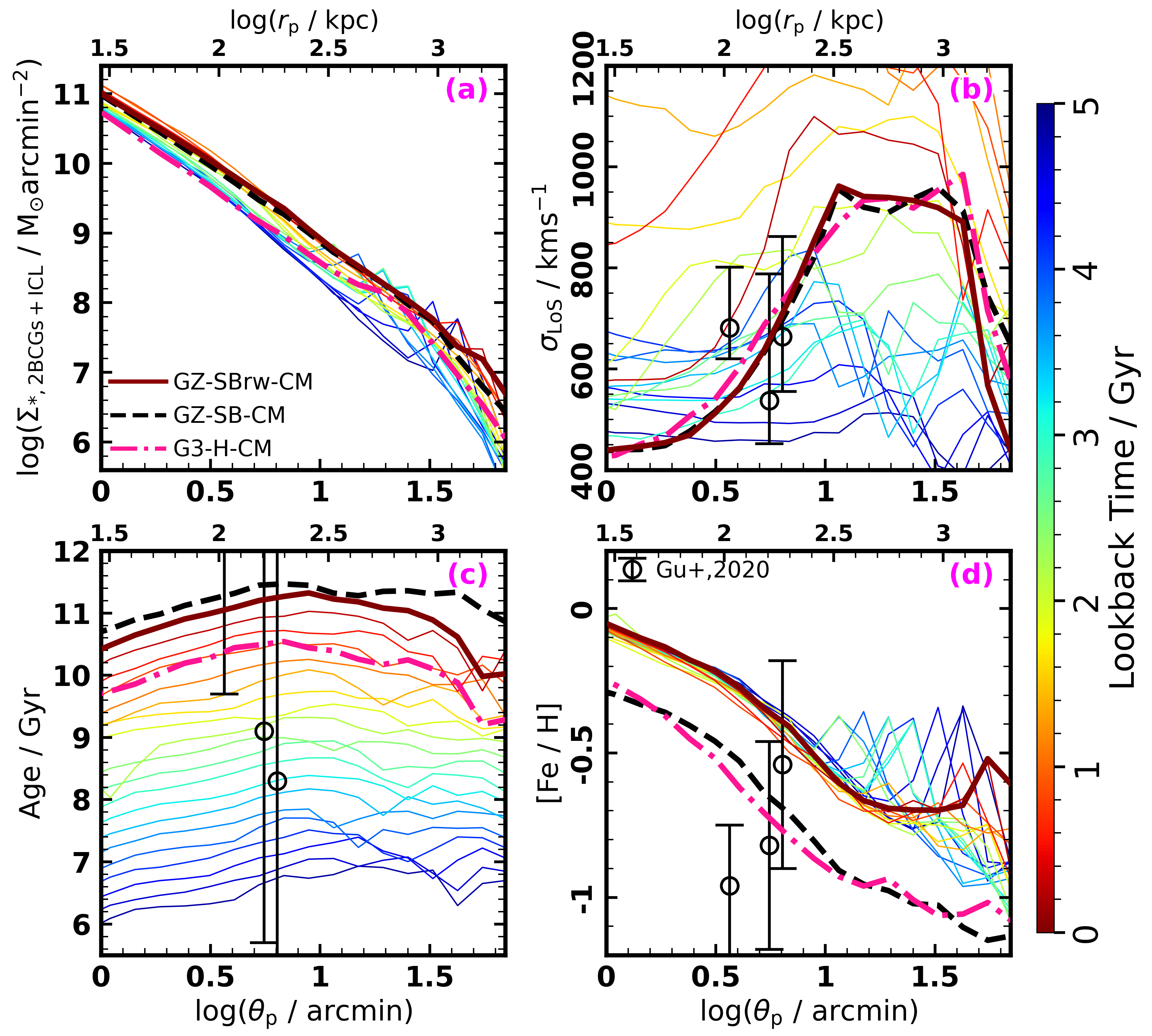}
    \caption{Property profiles of the 2BCGs+ICL component.
    The horizontal axis shows the angular distance to the nearest BCG, while the top axis indicates the corresponding projected distance.
    The four panels present different properties: (a) stellar surface density; (b) line-of-sight velocity dispersion; (c) stellar age; (d) stellar iron abundance.
    The solid lines represent simulation results at different redshifts (color-coded by lookback time), with the $z=0$ line (dark red) thickened for clarity.
    The black points with vertical error bars are observational results from \citet{Gu2020ApJ_ICL_Coma}.
    The black dashed and pink dash-dotted lines in each panel show simulation results at $z=0$ from GZ-SB-CM and G3-H-CM, respectively.
    }
    \label{fig:ICL_property_profile}
\end{figure*}

\citet{Gu2020ApJ_ICL_Coma} measured the properties of the ICL in three regions around the Coma cluster using MaNGA observations.
We compare their results with our simulations in \autoref{fig:ICL_property_profile}, which shows the evolutionary history of the 2BCGs+ICL property profiles as functions of the projected angular distance to the nearest BCG ($\theta_{\rm p}$).
Panel (a) presents the surface density profile $\Sigma_{*,\rm 2BCGs+ICL}$, while
panels (b)-(d) display the mass-weighted line-of-sight velocity dispersion $\sigma_{\rm LoS}$, stellar age, and iron abundance [Fe/H], respectively.
The thick dark red solid line in each panel represents the redshift-zero result from the GZ-SBrw-CM simulation, with the corresponding results for GZ-SB-CM and G3-H-CM shown as black dashed and pink dash-dotted lines, respectively.
The observational results are plotted as black circles with error bars.

As one can see, the surface density profile drops rapidly with increasing distance.
The normalizations are nearly identical for the models with AGN feedback (GZ-SBrw-CM and GZ-SB-CM), while G3-H-CM exhibits significantly lower profiles, differing by roughly half an order of magnitude. 
Given that the total ICL fraction varies by only $11.3\%$ among these models, we conclude that the surface density is more closely tied to the total stellar budget, whereas the ICL fraction is more sensitive to the dynamical evolution and merging history of the cluster.

For $\sigma_{\rm LoS}$, the observed values are $\sim 600 \ {\rm km}\cdot {\rm s}^{-1}$, which are broadly consistent with the simulation at the corresponding radii.
As a dynamical quantity, the differences among the models are minimal. 
The simulated $\sigma_{\rm LoS}$ profile remains relatively flat both near the two BCGs and in the region between $300-1000 \ {\rm kpc}$, with values of around $400$ and $900 \ {\rm km}\cdot {\rm s}^{-1}$, respectively, and shows a sharp rise over intermediate radii.

With regard to stellar age, the simulation results fall within the range of the observations, given the large uncertainties in the observational data.
All models show a profile that varies little with radius (within $1 \ {\rm Gyr}$).
The profiles from different models share similar shapes, with a difference of $\sim 1 \ {\rm Gyr}$ between them, which is likely related to their different star formation histories at early epochs, driven by the adopted feedback models.

The iron abundance profile of GZ-SBrw-CM matches the observational data point at the largest distance but significantly deviates from the two data points in the inner regions.
While GZ-SB-CM and G3-H-CM can reproduce the data point at the intermediate distance.
Moreover, none of the three simulation models can match the innermost data point.
All three ICL regions show iron abundances that are significantly below the observed value in the outskirts of the two BCGs, where the values are around $-0.4$ \citep{Gu2020ApJ_ICL_Coma}, which is consistent with the trends seen in our simulation profiles.
Furthermore, the iron abundance can differ by up to 0.3 dex between models, indicating the potential of using ICL metallicities to calibrate simulation models.

Generally, the simulation profiles at redshift zero match the observations, except for the observed data point closest to the BCGs, which corresponds to a special ICL region located near the midpoint of the two BCGs. 
Given the limited number of data points and the large observational uncertainties, future ICL observations are needed to effectively differentiate between galaxy formation models.

\subsection{Intracluster Filaments}
The dark matter in the outskirts of the cluster is asymmetric and forms structures such as intracluster filaments, which serve as the terminals of large-scale cosmic filaments feeding into the clusters \citep{Kuchner2020MNRAS_filaments, Kuchner2022MNRAS_filaments, Rost2021MNRAS_ThreeHundred_filament, Rost2024MNRAS_ThreeHundred_filament, Kotecha2022MNRAS_filament}.
\citet{HyeongHan2024NatAs_ICF} employed two different approaches, the matched-filter statistic and the shear-peak count statistic, to analyze the weak-lensing signal of the ICFs using Hyper Suprime-Cam imaging data of the Coma cluster, in the outskirt region spanning from 1 Mpc to 2.8 Mpc.
The matched-filter statistic measures the tangential shear signal correlated with a filament template convergence profile, providing the E-mode alignment strength of the lensing signal relative to the expected filament profile as a function of polar angle. 
The shear-peak count statistic, on the other hand, counts the shear-peak number density in pan-shaped region with an opening angle of $30^{\circ}$ in different directions, based on the mass map produced using a truncated Navarro–Frenk–White (NFW) profile aperture.
For both methods, a higher value along a given direction indicates a higher filamentary mass density or a stronger filament signal.
Using these two methods, they identified two robust ICFs along the north (N) and west (W) directions, as well as an ICF candidate along southeast (SE).
The corresponding significances for the N, W, and SE ICFs are $6.6 \sigma$, $3.6 \sigma$, and $4.3 \sigma$ for the matched-filter technique, and $3.1 \sigma$, $2.8 \sigma$, and $2.0 \sigma$ for the shear-peak count statistic, respectively.
Based on the matched-filter statistic, they obtained the normalization amplitude (related to the maximum surface mass density) at the filament ridge for the three ICFs, yielding $\kappa_0 = 0.0168_{-0.0023}^{+0.0024}$, $0.0188_{-0.0044}^{+0.0055}$, and $0.0080_{-0.0015}^{+0.0018}$, respectively.
Thus, the SE ICF is about a factor of two weaker the other two.
All of these ICFs are well aligned with the large-scale filaments previously identified by \citet{Malavasi2020A&A_spider} using galaxy overdensities. 

\begin{figure*}[htbp!]
    \centering
    \includegraphics[width=16cm]{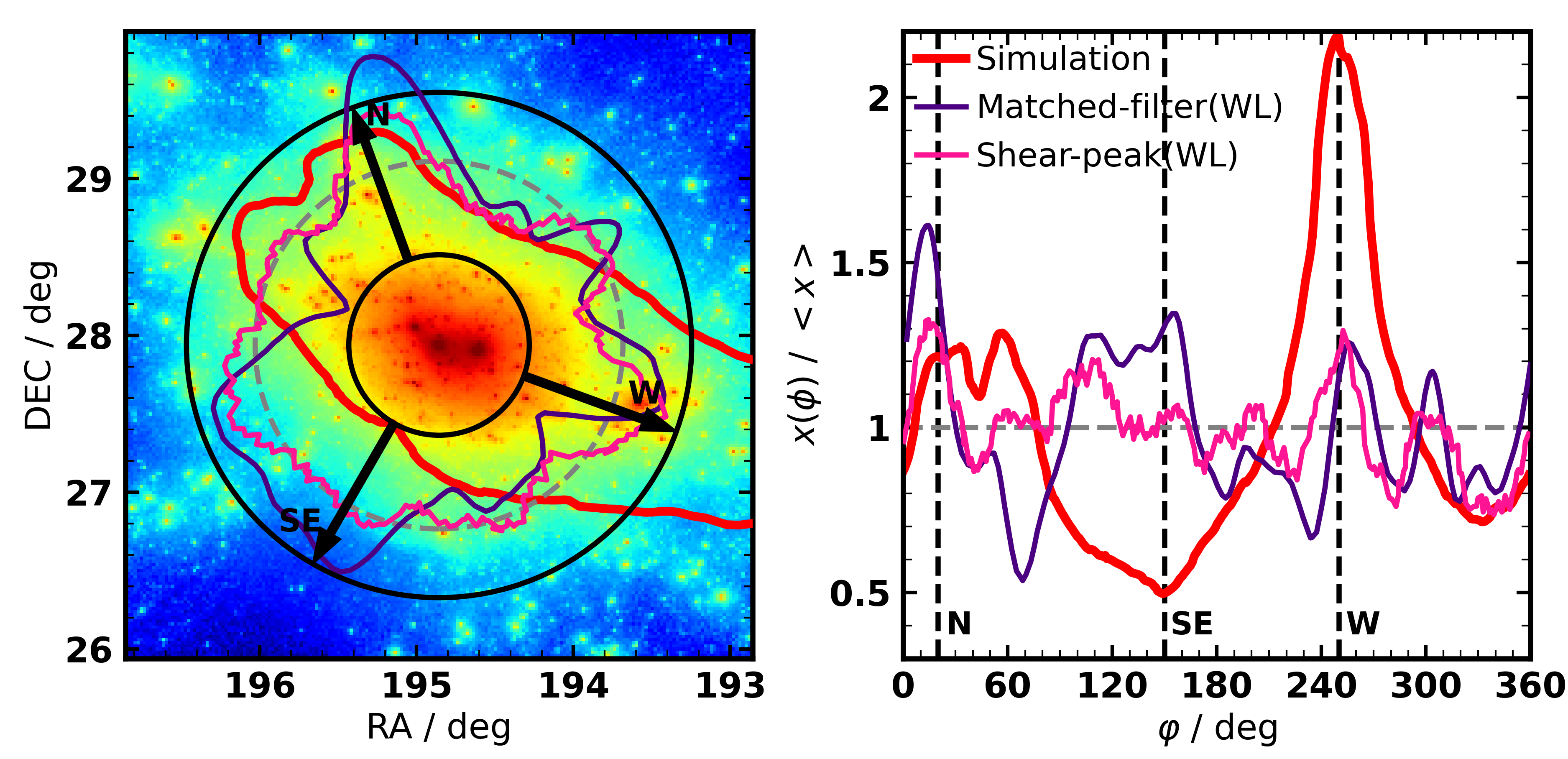}
    \caption{Intracluster filaments (ICF) in the simulation and observation.
    Left panel: 
    In polar coordinates centered on the cluster center, the purple and pink solid lines show the observational ICF signals along different directions from \citet{HyeongHan2024NatAs_ICF}, derived from the matched-filter and shear-peak count statistics, respectively (see the text for details).
    For comparison, the red solid line shows the surface mass densities from the simulation, measured over a sector angle of $30^{\circ}$ in each direction.
    Both the simulation and observational results are normalized by their respective azimuthal averages; the unit value is indicated by the gray dashed circle, which also marks the $R_{\rm 200c}$ radius of the cluster.
    The black arrows represent the positions of the three ICFs (N, W and SE) identified by the observation.
    The annulus between projected radii of $1 \ {\rm Mpc}$ and $2.8 \ {\rm Mpc}$ is delimited by the two black circles and corresponds to the region used for the calculations.
    The colored background map shows the surface mass density of all matter projected within a line-of-sight depth of $10 \ h^{-1}{\rm Mpc}$ centered on the simulated Coma cluster, with redder (bluer) colors indicating higher (lower) densities.
    Right panel: Direct comparison between simulation and observations as a function of position angle $\varphi$ measured from north to east ($0^{\circ}$ at north, increasing to $90^{\circ}$ at east).
    The solid lines are the same results as in the left panel.
    The black dashed vertical lines mark the directions of the three ICFs from the weak-lensing detection.
    }
    \label{fig:ICF}
\end{figure*}

To mimic the weak-lensing measurements, we compute the surface mass density of the simulated cluster along different directions within the same annular region between the projected radii of $1 \ {\rm Mpc}$ and $2.8 \ {\rm Mpc}$.
Each density value is calculated over a sector with an opening angle of $30^{\circ}$, projected within a depth of $10 \ h^{\rm -1}{\rm Mpc}$.
For a fair comparison, we normalize both the simulation results and the observational results from the two methods by their respective azimuthal averages. 
The solid red line and the solid purple and pink lines in \autoref{fig:ICF} represent the results of the simulation and the two observational methods, respectively.
The left panel presents a comparison in polar coordinates, while the right panel shows the same quantities as a function of position angle $\varphi$ measured from north to east. 
The positions of the three observed ICFs (N, W and SE) are indicated by the black arrows in the left panel and by the black dashed vertical lines in the right panel.
The background map in the left panel shows the surface mass density distribution in the simulation, and the annular region between the two black circles corresponds to the area used for the calculations.

The surface mass density of the simulated Coma cluster exhibits three peaks at $\varphi=33^{\circ},\ 57^{\circ}$, and $249^{\circ}$, indicating the presence of three ICFs along these directions, as can be seen in the colored background map.
Among them, the ICFs at $\varphi=33^{\circ}$ and $249^{\circ}$ are in good agreement with the observed N ($\varphi=20^{\circ}$) and W ($\varphi=250^{\circ}$) ICFs, although the simulation signal along the W direction is boosted by a substructure.
However, the observed SE ICF is not reproduced in our simulation.
It should be noted that its normalization amplitude is about a factor of two lower than that of the other two ICFs and was not considered a robust detection by \citet{HyeongHan2024NatAs_ICF}.
The additional ICF at $\varphi=57^{\circ}$ in the simulation lies close to the orientation of the northeast large-scale filament identified by \citet{Malavasi2020A&A_spider}, where the observational signals from the two methods also show a small peak.
Overall, the constrained simulation successfully reproduces the main ICF features of the real Coma cluster.

\subsection{Galaxy Groups}

The galaxy groups (including the subgroups within the Coma cluster) in the vicinity of the Coma cluster have been identified and studied in previous works.
\citet{Adami2005A&A_infalling_substructures} identified 17 groups (including the main cluster) and found that they are distributed along the directions toward the three neighboring clusters A 779, A 1367, and A 2199, suggesting that these groups may have been accreted from the surrounding large-scale structures.
\citet{Healy2021A&A_HI_Coma_substructures} applied the Dressler–Shectman (DS) test to a compiled redshift catalogue of the Coma cluster and identified 15 distinct groups with more than five members within the WCS HI data footprint.
More recently, \citet{Jimenez-Teja2025A&A_ICL_Coma} employed the DS+ method \citep{Benavides2023A&A_DSplus} on 2,157 galaxy members of the Coma cluster selected from the DESI redshift catalogs, resulting in 42 groups with at least six members, and found substructures similar to those reported in \citet{Adami2005A&A_infalling_substructures}.

\begin{figure*}[htbp!]
    \centering
    \includegraphics[width=16cm]{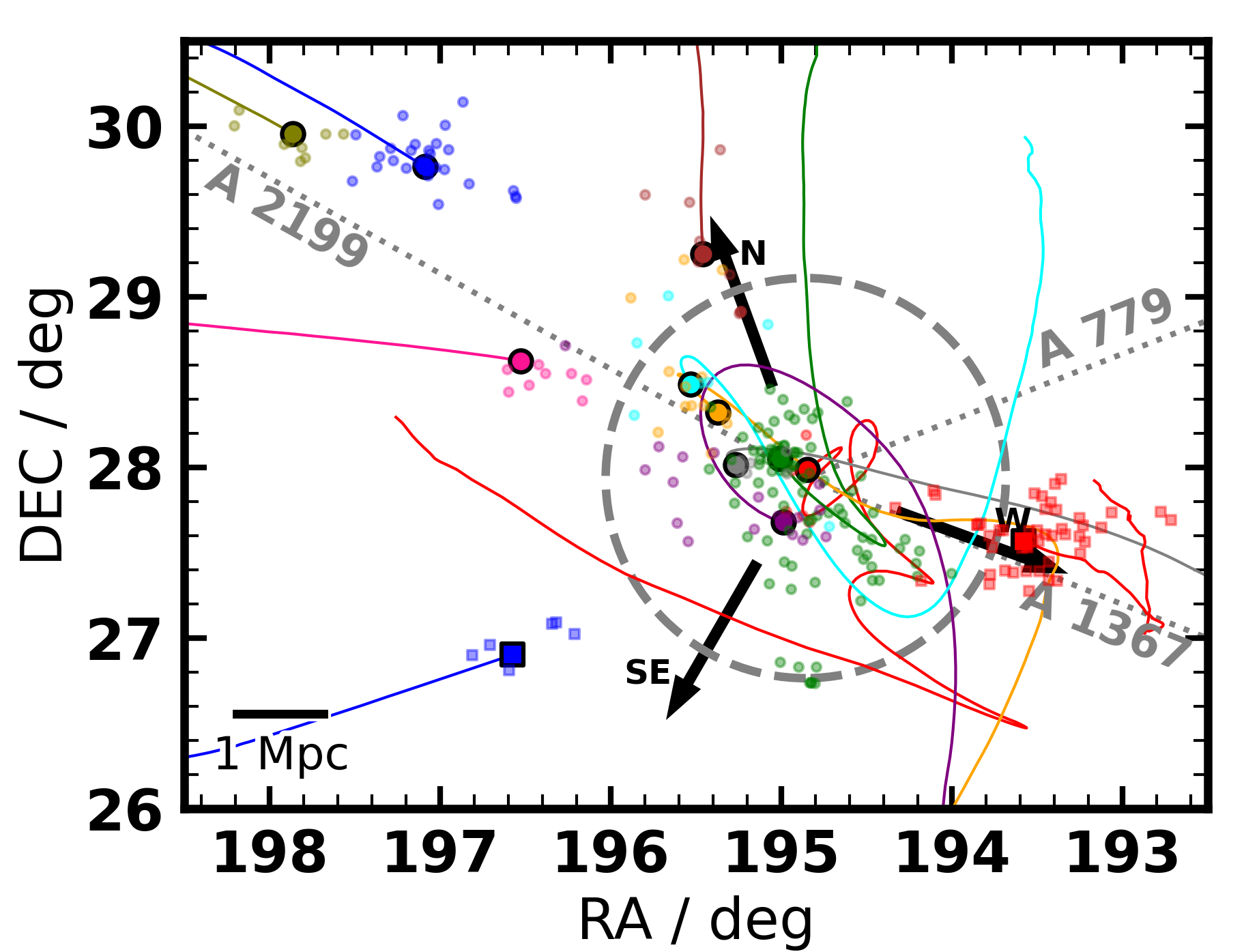}
    \caption{Spatial distribution of large galaxy groups in and surrounding the simulated Coma cluster at redshift zero, in J2000.0 coordinates.
    The displayed groups lie within a line-of-sight depth of $10 \ h^{-1}{\rm Mpc}$ centered on the Coma cluster, and are required to host at least six galaxies, each with a stellar mass $m_{*,\rm gal}>5\times 10^8 \ {\rm M}_{\odot}$ (corresponding to about 100 star particles).
    The larger markers with black outlines are the central galaxy of each group, while the smaller markers (same color and shape) are their member galaxies that satisfy the mass cut.
    For groups that are already within the Coma cluster, we show all their member galaxies at the snapshot immediately before they merge into the cluster, provided they still survive.
    The two subclusters containing the two BCGs are excluded.
    The historical trajectories of their central galaxies are shown by the solid lines with matching colors connected to them.
    The gray dotted lines indicate the directions toward the three nearby clusters in the real universe, while the three black arrows mark the positions of intracluster filaments identified by \citet{HyeongHan2024NatAs_ICF}.
    The gray dashed circle represents the $R_{\rm 200c}$ radius of the Coma cluster.
    }
    \label{fig:GalaxyGroups}
\end{figure*}

\autoref{fig:GalaxyGroups} shows the distribution of galaxy groups in the simulation at redshift zero on the sky plane.
In the simulation, we define galaxy groups based on the HBT subhalo catalogue, including both field subhalos (halos) and first-order subhalos (subgroups) within the Coma cluster.
Groups surrounding the Coma cluster that contain at least six galaxies (each satisfying $m_{*,\rm gal}>5\times 10^8 \ {\rm M}_{\odot}$) are shown with distinct markers, and their central galaxies highlighted by larger markers outlined in black.
The solid lines in matching colors trace their historical trajectories.
The directions of the three neighboring clusters are indicated by gray dotted lines \citep{Adami2005A&A_infalling_substructures}, while the positions of the ICFs from \citet{HyeongHan2024NatAs_ICF} are marked by three black arrows.

The overall distribution of the groups is consistent with observations: most are concentrated near the cluster core, while those in the outskirts tend to align with the ICFs and the directions toward nearby clusters.
Their distribution coincides with the ICF along the north and west directions, suggesting that they trace structures not only on large-scale filaments but also in the outer region of clusters.
Tracing their trajectories, we find that the groups in the outskirts have fallen in from directions primarily along the directions of A 2199 and A 1367, in agreement with observational suggestions.
In contrast, the groups close to the center have been accreted from various directions, including the north and south, where large-scale filaments are present \citep{Malavasi2020A&A_spider}.
The distinct origins of groups in different radial regimes reflect the anisotropic accretion pattern of the Coma cluster, which is primarily governed by the geometric configuration of the surrounding large-scale filaments.

\section{Predictions from Simulations} \label{Sec:Predictions}

A limitation of observations is that they cannot easily recover the full assembly history of a cluster, whereas constrained simulations can recover the history.
In the previous section, we compared our simulated Coma replica with the real Coma cluster and verified that it successfully reproduces many observed properties and structures both in and around Coma.
This gives us confidence to extract a formation history that is expected to be close to the true one.
In this section, we will present the predictions from the simulation regarding (i) the assembly history of the Coma cluster, (ii) the detailed merger history of the two BCGs, and (iii) the spatial distribution of the ICL properties.

\subsection{The Assembly History of the Coma Culster}

\begin{figure*}[htbp!]
    \centering
    \includegraphics[width=12cm]{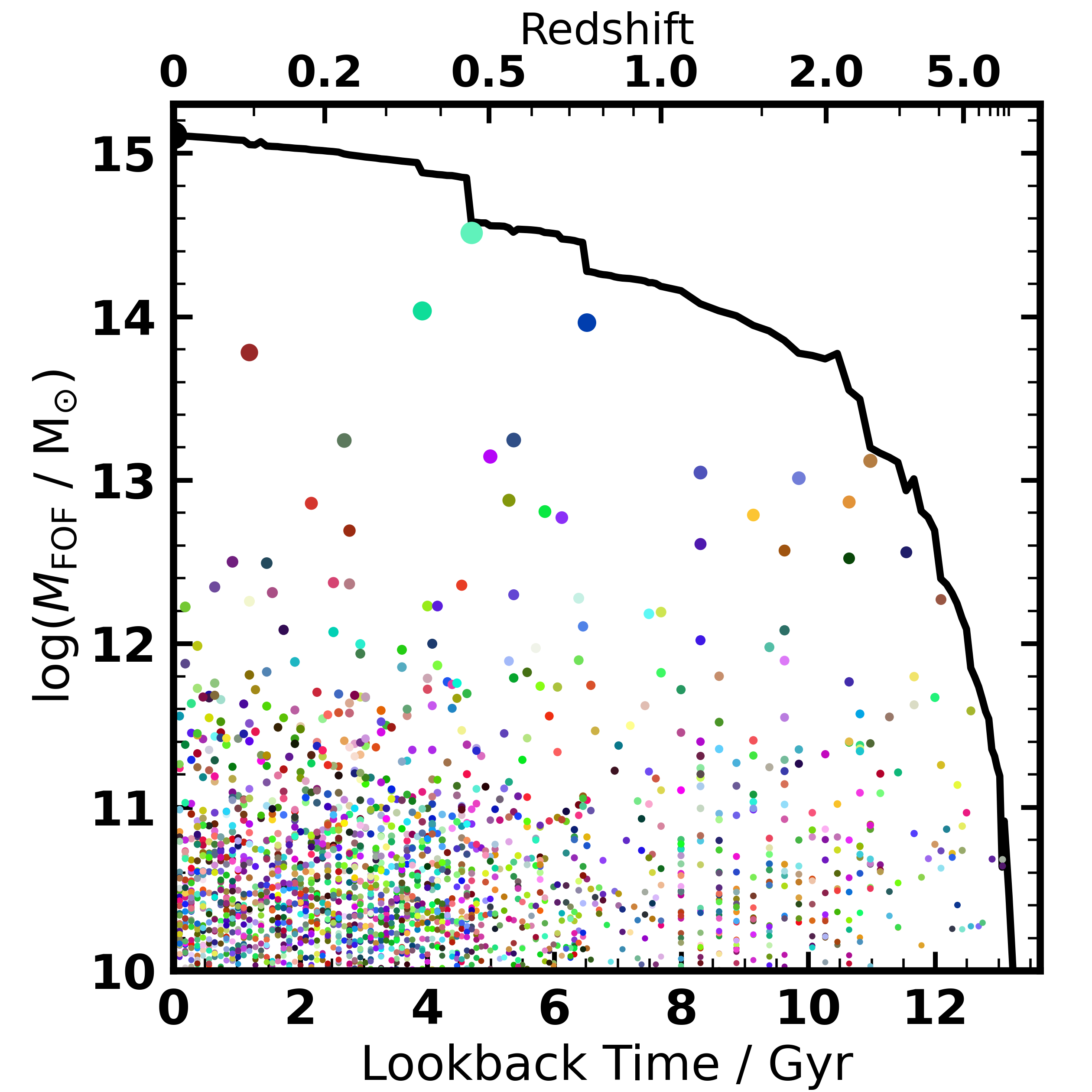}
    \caption{Assembly history of the Coma cluster.
    The black solid line shows the FOF halo mass of the Coma cluster as a function of lookback time.
    The colored points indicate the time and FOF halo mass of all halos at the snapshot immediately before their merger with Coma, with larger point sizes corresponding to higher halo masses.
    The corresponding redshifts are shown on the top axis.
    }
    \label{fig:ComaAssemblyHistory}
\end{figure*}

An advantage of constrained simulations is that they can provide the merger history of a cluster, as shown in previous works on the Virgo cluster \citep{Olchanski2018A&A_Virgo, Sorce2021MNRAS_CLONE}.
\autoref{fig:ComaAssemblyHistory} displays the overall assembly history of the FOF halo of the simulated Coma cluster.
The black line shows the evolution of its FOF halo mass as a function of lookback time.
Each point in the figure represents the mass of a halo at the snapshot immediately before it merges with the Coma cluster.

The Coma cluster reached the typical mass for a cluster of $10^{14} \ {\rm M}_{\odot}$ at $z=1.35$.
 Then it experienced two major mergers at $z=0.74$ and $z=0.45$, with halo mass ratios of 0.49 and 0.86, respectively.
The cluster had gained half of its total mass by $z\sim 0.45$, during the recent major merger between the host subclusters of the two BCGs observed at redshift zero.
As time progresses, mergers of smaller halos become much more frequent, particularly for halo masses below $10^{11} \ {\rm M}_{\odot}$ after $z\sim0.5$.
This result illustrates the complex assembly history of the Coma cluster.

\subsection{The Evolution Track of the Two BCGs}
In \autoref{fig:2BCGs_D_V} and \autoref{fig:2BCGsICL_2Dmap}, we have shown the history of the relative distance and velocity between the two BCGs as well as their trajectories, and compared them with observational data and other studies in \autoref{SubSec:BCGs}.
We find that the relative positions and velocities of the two BCGs at $z=0.021$ best match the observations.
Taking $z=0.021$ as the final state, we now describe their evolutionary paths.

Before their host clusters merge, BCG-W moves toward its final location along a position angle of $\sim 60^{\circ}$ (east of north), while BCG-E moves along a position angle of $\sim 220^{\circ}$ (i.e., south of west).
Their inclination angles relative to the sky plane are $\sim 50^{\circ}$ (receding from Earth) for BCG-W and $\sim 40^{\circ}$ (approaching Earth) for BCG-E respectively.
Their separation reached ${\rm R}_{\rm 200c} \sim2 \ {\rm Mpc}$ about $1.45 \ {\rm Gyr}$ ago, and then decreased to a minimum of $\sim 200 \ {\rm kpc}$ (an upper limit due to the limited snapshot cadence) about $0.64 \ {\rm Gyr}$ before the final state.
At the pericenter (yellow point in \autoref{fig:2BCGs_D_V}), their relative velocity reaches a maximum of $3728 \ {\rm km}/{\rm s}$.
In the final state (black point in \autoref{fig:2BCGs_D_V}), although their projected separation is only $0.3 \ {\rm Mpc}$, BCG-W is $1.4 \ {\rm Mpc}$ away from BCG-E along the line of sight and is close to apocenter.
Their relative velocity is $815 \ {\rm km}/{\rm s}$ at an angle of $23^{\circ}$ to the line-of-sight, indicating a motion that is nearly along the line-of-sight.

We also note that both BCGs experienced several mergers, including massive ones, along their path to convergence. 
For example, in the top row of \autoref{fig:2BCGsICL_2Dmap}, where the color map shows the distribution of all 2BCGs+ICL particles (initially identified at $z=0$) traced back to higher redshifts, the two BCGs are accompanied by galaxies that eventually merge with them (see middle panel for BCG-E and right panel for BCG-W).
These mergers further contribute to the final ICL content in the Coma cluster.

\subsection{The Properties of the Intracluster Light}

The overall ICL fraction is consistent with the observations, though it depends on the simulation model, as discussed in \autoref{SubSubSec:ICL_fraction}.
In addition to the enclosed ICL fraction profile, \autoref{fig:ICL_fraction_profile} also shows the fractional contributions of the 2BCGs (blue line) and the ICL (red line) as functions of projected angular distance in the right panel.
We find that the ICL dominates the stellar component from the outer regions of the 2BCGs out to a projected distance of about $150 \ {\rm kpc}$, where the enclosed ICL fraction (red solid line) peaks at a maximum value of $\sim30-40\%$.

\autoref{fig:ICL_property_profile} displays the evolutionary history of the 2BCGs+ICL profiles of surface density, line-of-sight velocity dispersion, stellar age, and iron metallicity.
The overall surface density profile (panel a) grows gradually with time, while its slope becomes slightly flatter.
The line-of-sight velocity dispersion (panel b) remains low, around $400-500 \ {\rm km}/{\rm s}$, until the cluster experiences significant accretion and mergers.
Subsequently, the profile becomes chaotic, especially in the central regions (around the two BCGs).
Eventually, the dispersion decreases near the center and peaks in the plateau between approximately 300 and 1000 kpc, with values around 900 km/s.
This high-dispersion ICL component mainly originates from stripped galaxies and preprocessed ICL, both of which are closely tied to the merger history, as illustrated by the phase space diagram of the ICL \citep{Jeon2026ApJ_NEWCLUSTER}.
Similar to the surface density, the stellar age profile (panel c) increases with time, while the location of its maximum shifts slightly inward.
At redshift zero, the maximum occurs at $\sim250 \ {\rm kpc}$ with an age of 11.3 Gyr.
The metallicity profile has evolved the least, especially in the central region; however, it is the most sensitive to the simulation model.
It drops rapidly with increasing projected radius, except in the outermost region, where it can be affected by ICL substructures stripped from galaxies.

\begin{figure*}[htbp!]
    \centering
    \includegraphics[width=16cm]{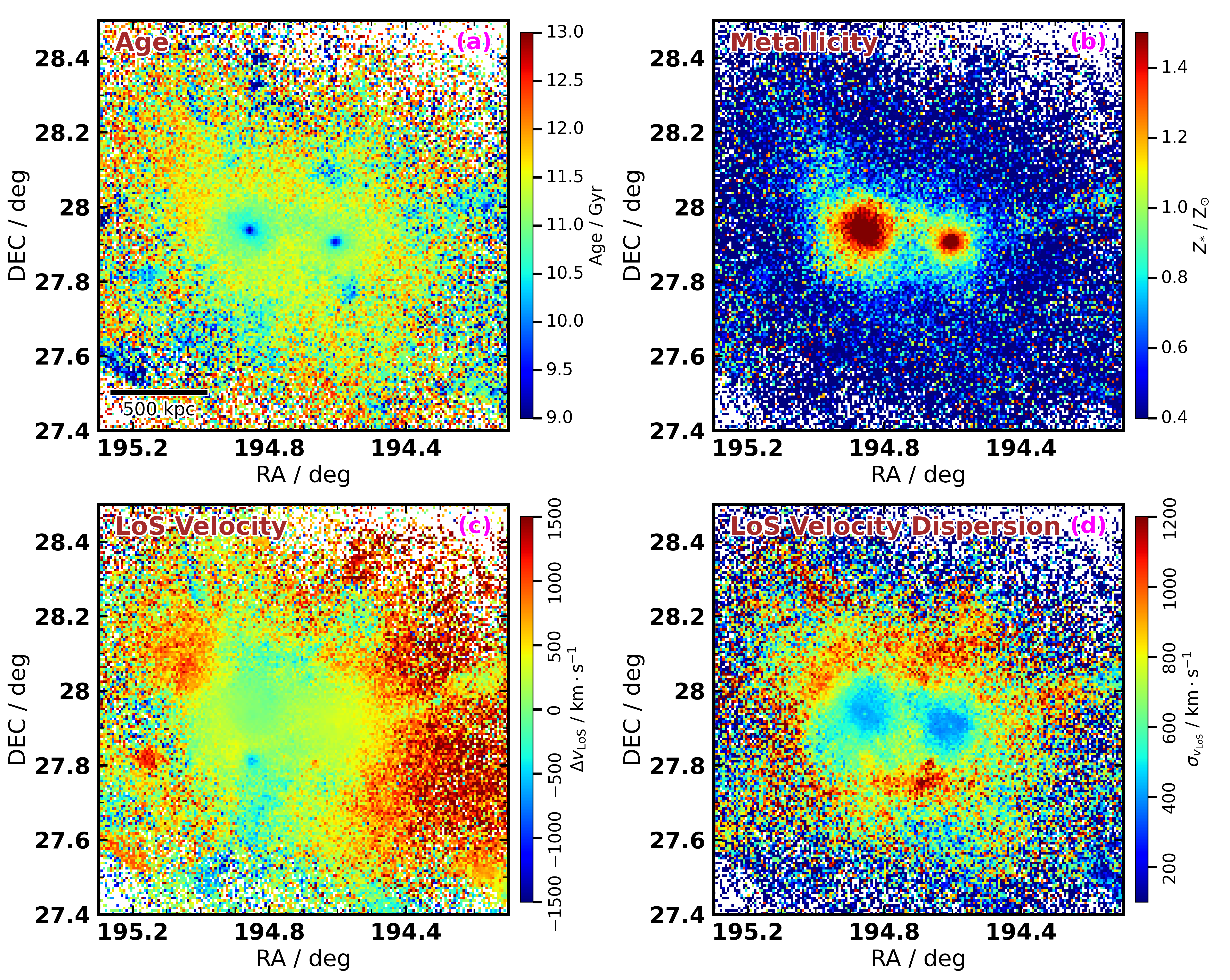}
    \caption{The property maps of the 2BCGs+ICL component.
    The maps show mass‑weighted properties computed from all 2BCGs+ICL particles at $z=0$, projected onto the sky plane within a line‑of‑sight depth of $10 \ {\rm Mpc}$.
    The mass-weighted properties are calculated using all 2BCGs+ICL particles at $z=0$, within a depth of $10 \ {\rm Mpc}$ along line of sight and projected to the sky plane.
    The four panels present, respectively: (a) stellar age, (b) metallicity $Z_{*}$, (c) line‑of‑sight velocity ($\Delta v_{\rm LoS}$, relative to the Coma center), and (d) line‑of‑sight velocity dispersion $\sigma_{v_{\rm LoS}}$.
    Each map is color‑coded according to the property value, as indicated by the color bar to the right of the panel.
    }
    \label{fig:2BCGsICL_property_map}
\end{figure*}

\autoref{fig:2BCGsICL_property_map} shows the distribution of the 2BCGs+ICL properties in the sky plane.
The four panels display mass-weighted properties: (a) stellar age, (b) metallicity $Z_{*}$, (c) line‑of‑sight velocity ($\Delta v_{\rm LoS}$, with the central velocity of the Coma cluster subtracted), and (d) line‑of‑sight velocity dispersion $\sigma_{v_{\rm LoS}}$.
The two BCGs host stars that are younger, more metal-rich, and have a lower velocity dispersion.
In contrast, the surrounding ICL tends to be older, much more metal-poor, and exhibits a higher velocity dispersion (see also the property profiles in \autoref{fig:ICL_property_profile}).
The properties of the outer ICL are anisotropic, with line-of-sight velocities preferring recession (positive $\Delta v_{\rm LoS}$).
The ICL also shows structures with high velocity dispersions, such as the westward region and the arc-shaped feature to the southeast, which are associated with stars stripped from galaxies that have passed through (or passed by) the cluster center.
These ICL features are the result of the history of the cluster center assembly and can be used to recover the merger history of the BCGs \citep[e.g., ][]{Gu2013ApJ_Minor_Mergers}.

\section{Summary and discussion} \label{Sec:Summary}
In this work, we use constrained hydrodynamic simulations of the Coma cluster from the ELUCID project to investigate its substructures and assembly history.
Our simulations accurately recover the cluster's global properties, including its position, virial mass, surrounding large-scale filaments, ICF, ICL properties, and the distribution of surrounding galaxy groups, demonstrating the power of constrained simulations for studying individual galaxy clusters.
Our main findings are summarized as follows:

\begin{itemize}[label=--, leftmargin=1.5em]
    \item \textbf{Global properties and large-scale environment.}
    The simulated Coma cluster is located at $\rm RA=194.86^{\circ}, \ DEC=27.94^{\circ}$, with a distance of $\rm D=99.6 \ {\rm Mpc}$, in excellent agreement with the observed values. 
    Its virial mass $M_{\rm 200c}=7.3\times10^{14} \ h^{-1}{\rm M}_\odot$ and radius $R_{\rm 200c} = 1.47 \ h^{-1}{\rm Mpc}$ are consistent with weak-lensing and X-ray measurements.
    The simulation also reproduces the large-scale filaments around Coma, particularly the north-east and west filaments identified by previous observational studies.
    
    \item \textbf{The two brightest cluster galaxies.}
    We find that the relative projected angular distance and line-of-sight velocity of the two BCGs at $z=0.021$ best match the observations, suggesting a small time delay of $\sim 0.28\ {\rm Gyr}$ between the simulation and the real cluster.
    The BCG trajectories indicate a merger close to the line-of-sight direction, in contrast to the plane-of-sky convergence scenario proposed by some observational studies. 
    BCG-W moves toward the final position of Coma with a position angle of $\sim60^{\circ}$ east of north, while BCG-E moves with $\sim 220^{\circ}$.
    Their pericentric passage occurred $\sim 0.64 \ {\rm Gyr}$ before the final state, after which they receded to apocenter with a line-of-sight separation of $\sim 1.4 \ {\rm Mpc}$.
    
    \item \textbf{The intracluster light.}
    Using a combined HBT+SKID method to define the 2BCGs+ICL component, we obtain a total ICL fraction of $18.3\%$ for our fiducial model, with the enclosed fraction reaching $21.2\%$ at $1.33 \ {\rm Mpc}$ and $20.0\%$ at $1.59 \ {\rm Mpc}$, consistent with recent observations. 
    The ICL fraction ranges from $12.2\%$ to $23.5\%$ across the different models, all within the range observed for merging clusters.
    The ICL profiles of velocity dispersion, stellar age, and iron abundance generally agree with MaNGA measurements, except for a higher than observed iron abundance for all the models for the most central observed radius.
    The ICL exhibits a clumpy east-west elongated morphology in the core, which is closely linked to the merger history of the two BCGs.
    
    \item \textbf{The intracluster filaments.}
    The simulation reproduces the north and west ICFs detected by weak lensing, with peaks at $\varphi=33^{\circ}$ and $249^{\circ}$, in good agreement with the observed directions ($\varphi=20^{\circ}$ and $250^{\circ}$).
    The observed southeast ICF is absent in our simulation, consistent with its low detection significance.
    An additional ICF at $\varphi =57^{\circ}$ lies close to the orientation of the north-east large-scale filament identified in the observations.
    Overall, the constrained simulation successfully recovers the main ICF features of the real Coma cluster.

    \item \textbf{Galaxy groups.}
    The spatial distribution of the surrounding galaxy groups is consistent with observations, with most groups concentrated near the cluster core and those in the outskirts aligned with the ICFs and directions towards the nearby A 2199 and A 1367 clusters.
    Tracing their trajectories, we find that groups in the outskirts have fallen in primarily along the directions of A 2199 and A 1367, while groups closer to the center have been accreted from various directions.
    
    \item \textbf{Assembly history.}
    We predict that the Coma cluster reached the typical cluster mass of $10^{14} \ {\rm M}_{\odot}$ at $z=1.35$, and experienced two major mergers at $z=0.74$ and $0.45$, with halo mass ratios of 0.49 and 0.86, respectively.
    After $z\sim0.5$, mergers of smaller halos (below $10^{11} \ {\rm M}_{\odot}$) became much more frequent.
    
\end{itemize}

The ELUCID reconstruction project is designed to recover the mass density field and the formation history of large-scale ($\gtrsim {\rm Mpc}$) structures in the SDSS region. In the reconstruction volume, all structures are handled in the same way. Therefore, it may appear surprising that substructures (typically $<{\rm Mpc}$) inside the Coma galaxy cluster can also be reproduced accurately. We stress that the substructures examined here, such as ICFs, ICL, and galaxy groups, are tightly connected to the surrounding large-scale environment, which is robustly recovered by the reconstruction. As discussed above, even the two BCGs, the most strongly non-linear system considered in this work, show a close association with the large-scale structures. This likely explains why our constrained simulations can reliably reproduce these highly non-linear substructures.

We mainly report the simulation results at $z=0$. This choice is based on the fact that the reconstruction was intended to reproduce present-day large-scale structure, including the Coma cluster (even though it is observed at a slightly higher redshift). Our analysis indicates that the three substructure components—ICL, ICFs, and galaxy groups—change only gradually near redshift zero, so the conclusions do not depend sensitively on the exact redshift adopted. In contrast, the dynamical state of the two-BCG system varies rapidly. We therefore trace its evolution and find that the simulations agree best with the observations at $z=0.021$.

In conclusion, our results demonstrate that constrained simulations are a powerful tool for bridging the gap between observations and the unobservable assembly histories of individual galaxy clusters. They provide a robust framework for interpreting the rich substructures of the Coma cluster and for calibrating galaxy formation models. 
Future ICL observations with improved spatial coverage and sensitivity, combined with higher-resolution simulations, will be essential for further constraining the physical processes governing the formation and evolution of intracluster light and its connection to the cluster merger history. 
In future work, we plan to investigate the formation of various cluster components, including the ICL, UDGs, and the environmental effects on galaxies within the cluster.

\section*{Acknowledgements}

This work is supported by the National Natural Science Foundation of China (NSFC, Nos. 12595312, 12192224, 12595311, 12133006). HYW thanks the support of CAS Project for Young Scientists in Basic Research, Grant No. YSBR-062, the New Cornerstone Science Foundation through the XPLORER PRIZE, and the China Manned Space Program
with grant no. CMS-CSST-2025-A04. 
The work is supported by the Supercomputer Center of University of Science and Technology of China.

\bibliographystyle{aasjournalv7}
\bibliography{reference}

\end{document}